\documentclass[%
 reprint,
 amsmath,amssymb,
 aps,
]{revtex4-2}

\usepackage{graphicx}
\usepackage{dcolumn}
\usepackage{bm}
\usepackage{hyperref}
\usepackage{amsmath}
\usepackage{amssymb}
\usepackage{graphicx}
\usepackage{dcolumn}
\usepackage{bm}
\usepackage[inline]{enumitem}
\usepackage{xcolor}
\usepackage{dsfont}
\usepackage{verbatim}
\usepackage{mathtools, xcolor,ytableau, amsfonts,tikz}

\newcommand{\vev}[1]{\langle #1 \rangle}
\NewDocumentEnvironment{alignb}{b}{%
  \begin{align*}
  \refstepcounter{equation} #1 \tag{\theequation}
  \end{align*}}

\begin{document}

\title{The Baryon Junction and String Interactions}

\author{Zohar Komargodski}
\email{zkomargodski@scgp.stonybrook.edu}
\affiliation{Simons Center for Geometry and Physics, SUNY, Stony Brook, NY 11794, USA}
\affiliation{C. N. Yang Institute for Theoretical Physics, Stony Brook University, Stony Brook, NY 11794, USA}

\author{Siwei Zhong}
 \email{siwei.zhong@stonybrook.edu}
\affiliation{Simons Center for Geometry and Physics, SUNY, Stony Brook, NY 11794, USA}
\affiliation{C. N. Yang Institute for Theoretical Physics, Stony Brook University, Stony Brook, NY 11794, USA}

\date{\today}

\begin{abstract}
We study junctions between confining strings. We show that the effective theory of such junctions is very predictive with only one new parameter, the junction's mass, controlling the first couple of terms in the expansion in the system size. By open-closed duality these considerations about the baryon junction map to interaction vertices of closed strings. Therefore, we calculate the interaction vertices of closed strings in theories such as Yang-Mills theory. We find some surprising selection rules for string interactions in 3+1 dimensions. Requiring perturbative stability and that the string coupling is weak, we suggest constraints on the junction's mass. 
\end{abstract}

\maketitle

\section{Introduction}

Many theories admit unbreakable string-like excitations of nonzero tension $T=l_s^{-2}$. The quintessential example is the Abrikosov string in superconductors~\cite{Abrikosov1957zh} and the closely related Nielsen-Olesen strings in the Abelian Higgs model~\cite{Nielsen:1973cs}. 
The string stability is a consequence of unbroken 1-form symmetries in the system. In the Abelian Higgs model, it is the magnetic $U(1)$ 1-form symmetry and magnetic flux is confined in the Abrikosov-Nielsen-Olesen string. Another model is the $SU(N)$ Yang-Mills theory with $\mathbb{Z}_N$ electric 1-form symmetry, where string-like excitations confine electric flux. 

The examples above are gapped theories in the bulk. In the presence of a long confining string, the only low-energy modes are the string fluctuations (phonons). The effective theories of such fluctuations are well-studied, both for closed and open strings, see for instance~\cite{Polchinski:1991ax,luscher2004string,Drummond:2004yp,aharony2009effective,aharony2010corrections,aharony2011effective,aharony2012effective, dubovsky2012effective,Meineri:2012uav, aharony2013effective,Brambilla:2014eaa,Hellerman:2014cba,Hellerman:2017upi,EliasMiro:2019kyf,Caselle:2021eir,EliasMiro:2021nul}. Some results are nicely reviewed in~\cite{Brandt:2016xsp}. There is also extensive literature on simulations of the confining string and the comparison with theoretical predictions~\cite{Teper:2009uf,Brandt:2009tc,Athenodorou:2010cs,Billo:2011fd,Athenodorou:2011rx,Brandt:2013eua,Caselle:2015tza,Athenodorou:2016kpd,Brandt:2017yzw,Athenodorou:2020ani,Caristo:2021tbk,Luo:2023cjv} (many more references on this subject can be found therein). Many papers also explored the subject of quantizing the string with dynamical endpoints and making contact with Regge physics, e.g.~\cite{Zahn:2013yma,Sonnenschein:2014jwa,Helle,Hellerman:2016hnf,Sever:2017ylk,Sonnenschein:2018aqf,Sonnenschein:2018fph, hooft2004minimal,future}. 

In the open string EFT, a Dirichlet boundary condition physically represents external static particles on which the string can end, and a Neumann boundary condition models branes within which the string endpoint can roam freely. In gauge theories, the point particles on which the strings can end are just quarks.
One physical example of the Neumann condition is in Yang-Mills theories when the flux tube ends on the ``Janus'' interface that is created by changing the theta angle $\theta\to \theta+2\pi$. The dynamics of these interfaces and some aspects of the strings ending on them is discussed in~\cite{Gaiotto:2017tne}.

In theories with $\mathbb{Z}_N$ 1-form symmetry,
an interesting string configuration is the ``baryon'' where $N$ strings are tied at a vertex point. The string-and-junction configuration was first discussed in \cite{Artru:1974zn, Rossi:1977cy}. This configuration plays an important role in particle scattering, as it tracks baryon number, see for instance~\cite{Kharzeev:1996sq, Frenklakh:2023pwy, Frenklakh:2024mgu,altmann2024string} and is also essential toward understanding exotic hadrons, e.g.~\cite{Rossi:2016szw, PhysRevD.95.034011, Karliner:2020vsi}. In this paper, we will focus on confining theories with $\mathbb{Z}_3$ one-form symmetry in a $(d+1)$-dimensional spacetime. Our analysis is independent of the UV physics and bears easy generalization to other string-vertex configurations. Particularly, our study applies to the confined phase of $SU(3)$ Yang-Mills theory in $d=2,3$. The baryon vertex is clearly observed in simulations \cite{Takahashi:2000te, Takahashi:2002bw, Alexandrou:2002sn, Bissey:2005sk, Bissey:2006bz} where our predictions should be testable.

\begin{figure}[h!]
\centering
\includegraphics[width=0.5\textwidth ]{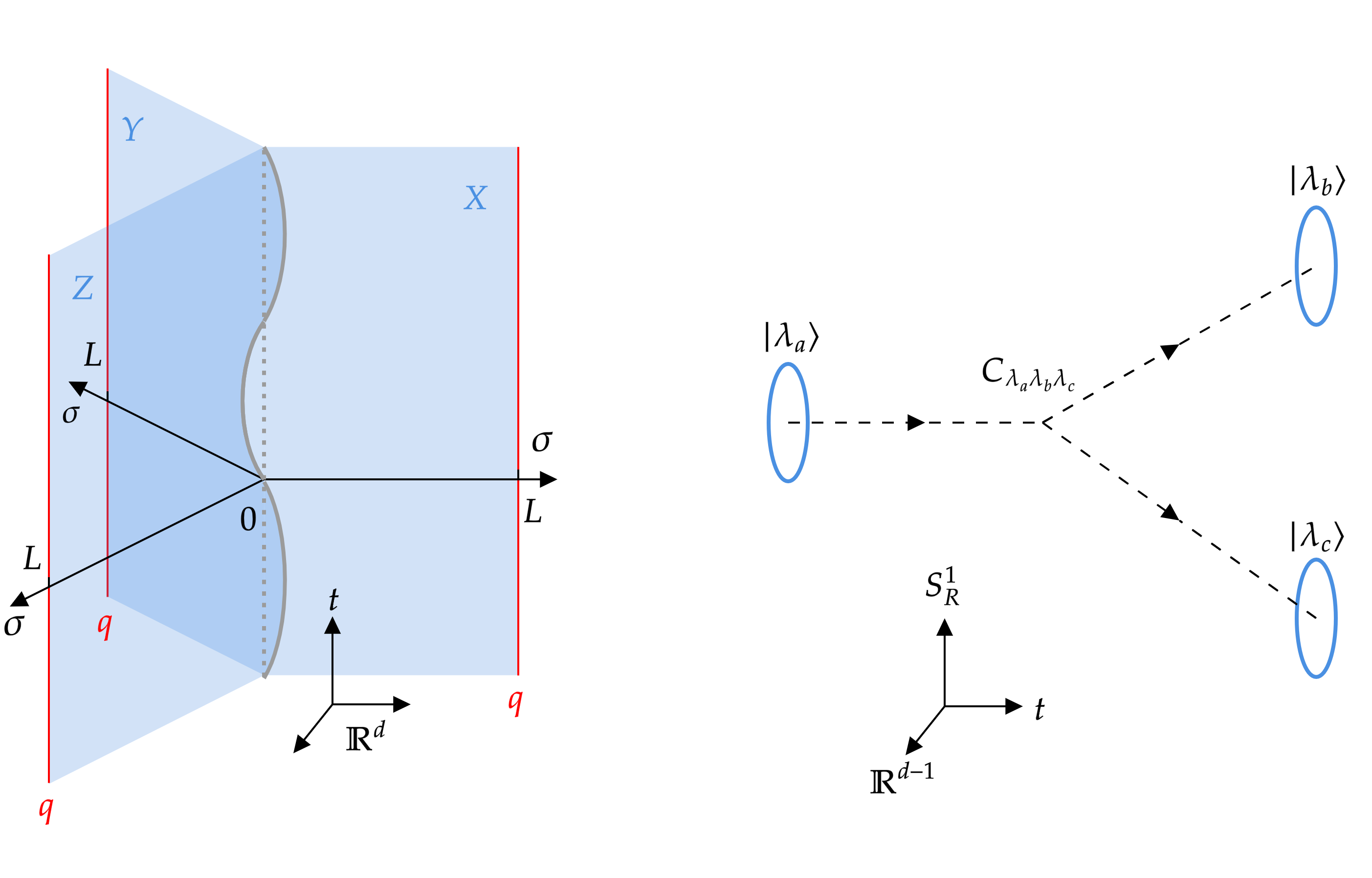}
  \caption{The ``Baryon'' Configuration. In the open channel (Left), three strings are tied at the Fermat point at $\sigma=0$ and end at the Dirichlet boundaries (quarks) at $\sigma=L$. In the closed channel (Right), three closed string states $\lambda_{a,b,c}$ have an interaction vertex  $C_{\lambda_a\lambda_b\lambda_c}$.}
  \label{pic_setup}
\end{figure}

We investigate the ``baryon'' from two perspectives. In the open channel, we consider three static quarks positioned at the vertices of a triangle. In this channel, confining strings meet at the Fermat point as in figure \ref{pic_setup}, and the time direction is perpendicular to the triangle. Equivalently, we can perform a double Wick rotation and define the closed channel, in which the time direction lies on the plane. The vertex is then interpreted as an interaction vertex where three strings meet.  We can call such a vertex a D-instanton as it is localized in time.

Let $L$ be the length of confining strings, we denote an EFT operator to be of order $k$ if it scales as $O(1/L^k)$ in the action and contributes $O(1/L^{k+1})$ corrections to the spectrum. We argue that the EFT is uniquely determined by $M$ and $l_s$ up to order 2.
Up to order $-1$ the physics is classical.  
The fact that the ``baryon'' configuration only breaks  Poincaré symmetry spontaneously imposes strong constraints on the EFT, which is why the EFT expansion is robust with no new parameters up to order 2.  The two parameters $M$ and $l_s$ can be obtained (for instance) by measuring the ground state energy of the junction, extending \cite{jahn2004baryons} (and related calculations in~\cite{Pfeuffer:2008mz}) 
\begin{equation}
\label{eq_ground state energy}
E_{\text{gs}}=\frac{3L}{l_s^2}+M-\frac{(d-2)\pi }{16L}-\frac{(d+2)\pi Ml_s^2}{144 L^2}+O(1/L^3)
\end{equation}
Note that in \eqref{eq_ground state energy} the static quark masses $m_{q}$ have been subtracted. The static quark masses $m_q$ are counter-terms on the Dirichlet boundaries corresponding to the external quarks, and hence, they are scheme-dependent. In any specific implementation they can be measured from the quark-antiquark pair and then subtracted to obtain~\eqref{eq_ground state energy}, which is scheme-independent. 

In this paper, we will explicitly demonstrate the implications of open-closed duality up to order 1. We will see that the string interaction vertices $C_{\lambda_a\lambda_b\lambda_c}$ can be obtained, in some cases completely unambiguously, just from consistency. We find that the coupling constant among generic closed string states of strings of length $2\pi R$ is 
\begin{equation} \label{couplingco}C_{\lambda_a\lambda_b\lambda_c}\sim e^{-2\pi MR}~.\end{equation} 
We see that if $M>0$ long strings are weakly coupled, and strongly coupled otherwise. We will also show that on top of the strong coupling catastrophe, there is a classical perturbative instability for sufficiently large negative $M$. 
With the same framework, many extensions of the calculations we do are possible, and we comment on some of the open problems towards the end.

\section{Review of Effective String Theories}

Here we review the effective field theory of a long confining string, which physically describes small fluctuations of the string. In the low-energy limit, we assume all massive modes have been integrated out, leaving an effective action of massless modes. A long string configuration spontaneously breaks spacetime Poincaré symmetry as $ISO(1,d)\to ISO(1,1)\times SO(d-1)$, leading to $(d-1)$ Nambu-Goldstone Bosons (NGBs) \cite{low2002spontaneously} as the only massless degrees of freedom in the string interior. 

To be concrete, we study a closed string wrapped on $\mathbb{S}^1_R$, whose worldsheet embedding in spacetime is $X_\mu(\Sigma)$ and $\Sigma_{\alpha}=(t,\sigma)$. The effective action is constrained by Poincaré symmetry and diffeomorphism invariance on the worldsheet. An important action compatible with these requirements is the Nambu-Goto action \cite{nambu1970duality, Goto:1971ce}: 
\begin{equation}
\label{eq_Nambu-Goto Action}
S_{\text{NG}} = - \frac{1}{l_s^2}\int dt d\sigma\sqrt{-\det \partial_\alpha X^\mu \partial_\beta X_\mu}~. 
\end{equation}
where we chose a mostly negative signature. Without loss of generality, we choose a static gauge where $X_0=t$, $X_1=\sigma \in \mathbb{S}_R^{1}$, and $X_i=l_sx_i(t,\sigma)$ for $2 \leq i\leq d$. The Nambu-Goto action admits an expansion in transverse fluctuations $x_i$, and in this gauge it reads:
\begin{multline}
\label{eq_Nambu-Goto Expansion}
S_{\text{NG}} =\int dt d\sigma \left\{-\frac{1}{l_s^2}+\frac{1}{2}\left[(\partial_t x_i)^2-(\partial_\sigma x_i)^2\right]\right.\\
\left.+\frac{l_s^2}{8}\left[(\partial_t x_i-\partial_\sigma x_i)^2(\partial_t x_{{i}'}+\partial_\sigma x_{{i}'})^2\right]\right\}+O(\frac{1}{R^4}).
\end{multline}
Equation~\eqref{eq_Nambu-Goto Expansion} is the unique effective action of NGBs $x_i$ to order 2  \cite{aharony2009effective,aharony2012effective, aharony2013effective}. There are corrections to~\eqref{eq_Nambu-Goto Action} and \eqref{eq_Nambu-Goto Expansion} with new Wilson coefficients, which starts at order 6 for $d=2$ and order 4 for $d\geq 3$ -- we will not be concerned with such high-order contributions here.

From the order $-2$ term in \eqref{eq_Nambu-Goto Expansion} we interpret $T=l_s^{-2}$ as the classical string tension. The order $0$ term consists of $(d-1)$ free NGBs, whose modes are left- or right-moving. The left and right mode occupation numbers $n^{\text{L,R}}_\lambda\in \mathbb{N}$ determine the order 0 energy level of a closed string state $\lambda$: 
\begin{equation}
\label{eq_closed string energy}
\begin{aligned}
E^{\text{c}}_\lambda=&\frac{2\pi R}{l_s^2}-\frac{(d-1)}{12R}+\frac{n^\text{L}_\lambda+n^\text{R}_\lambda}{R}+O(1/R^3)~.
    \end{aligned}
\end{equation}
The order $2$ term is a $T\Bar{T}$-deformation of the free action and hence preserves integrability~\cite{Caselle:2013dra,Dubovsky:2014fma,Chen:2018keo}, and it leads to $O(1/R^3)$ energy corrections in~\eqref{eq_closed string energy}. For later convenience, we denote the ground state by $\mathbf{0}$ -- it has  $n^\text{L}_{\mathbf{0}}=n^\text{R}_{\mathbf{0}}=0$ and the lowest-lying non-chiral $O(d-1)$ symmetric state is denoted by $\mathbf{1}$ -- it has  $n^\text{L}_{\mathbf{1}}=n^\text{R}_{\mathbf{1}}=1$.

The analysis is similar for open strings, except we need to consider boundary conditions and boundary operators. We take an open string of length $L$. At order 0, two canonical choices for the boundary condition at $\sigma=0$ are 
\begin{itemize} 
\item {Dirichlet}: $\partial_t x_i=0$. Among all boundary operators we note that $\int_{\sigma=0}dt (\partial_\sigma x_i)^2$ is forbidden by the spontanously broken Poincaré symmetry \cite{luscher2004string, aharony2011effective}. The leading operator is $\int_{\sigma=0} d t (\partial_t \partial_\sigma x_i)^2$ and it is of order 3. It represents the moment of inertia of the boundary endpoint.  

\item {Nuemann}: $\partial_\sigma x_i=0$. The leading boundary operator is $\int_{\sigma=0}dt (\partial_t x_i)^2$ and it is of order 1. The Poincaré symmetry requires this operator to be accompanied by an order $-1$ constant term, such that
\begin{equation}
\quad \quad S_\text{N}=-{M}'\int_{\sigma=0}dt\left[1-\frac{l_s^2}{2}(\partial_t x_i)^2\right]+O\left(1/L^3\right)
\end{equation}
We interpret ${M}'$ as the classical mass of the endpoint, which is free to roam on the brane at the boundary. The next order operators, such as $\int_{\sigma=0}dt [(\partial_t x_i)^2]^2$, are order 3. 
\end{itemize}

Finally, it is important to remark that the EFT expansion means that the frequency cannot exceed $l_s^{-1}$, otherwise the expansion in the inverse string size breaks down.

\subsection{The Open-Closed Duality for a ``Meson"}
\label{sec_meson}
As a warm-up, we review the ``meson", which is the configuration of an open string connecting a static quark-antiquark pair ($q\bar q $ pair). We denote the string spatial length to be $L$ and let boundary conditions be Dirichlet at $\sigma=0$ and $\sigma=L$. Furthermore, we compactify the time direction on $\mathbb{S}^1_R$ so as to put the open string at finite temperature. From \eqref{eq_Nambu-Goto Expansion} we learn that up to order 1 the unique meson action reads $S_\text{m}=S_\text{m}^{(-2)}+S_\text{m}^{(0)}+O(1/L^2)$, where the order $-2$ constant $S_\text{m}^{(-2)}=\frac{2\pi R L}{l_s^2}$ and the order 0 fluctuations read
\begin{equation}
\begin{aligned}
S_{\text{m}}^{(0)}
=&\frac{1}{2}\int d \tau d \sigma \left[(\partial_\tau x_i)^2+(\partial_\sigma x_i)^2\right]
\end{aligned}
\end{equation}
In the long string limit, higher operators are suppressed. For the meson case, the EFT partition function can be obtained as
\begin{equation}
\label{eq_meson partition function}
\begin{aligned}
\mathcal{Z}_{\text{m}}=&e^{-S_{\text{m}}^{(-2)}}\int \mathcal{D}x_i e^{-S_{\text{m}}^{(0)}}\left[1+O\left(1/L^2\right)\right]\\
=&\frac{e^{-\mu L}}{\left[\eta(q)\right]^{(d-1)}}\left[1+O\left(1/L^2\right)\right]
    \end{aligned}
\end{equation}
where the modular parameter $q\equiv e^{-\frac{2\pi^2 R}{L}}$ and $\eta(q)$ is the Dedekind eta function. $e^{-\mu L}$ represents the contribution from the classical energy of the string, where $\mu\equiv\frac{2\pi R}{l_s^2}$.
\begin{figure}[h!]
\centering
\includegraphics[width=0.5\textwidth ]{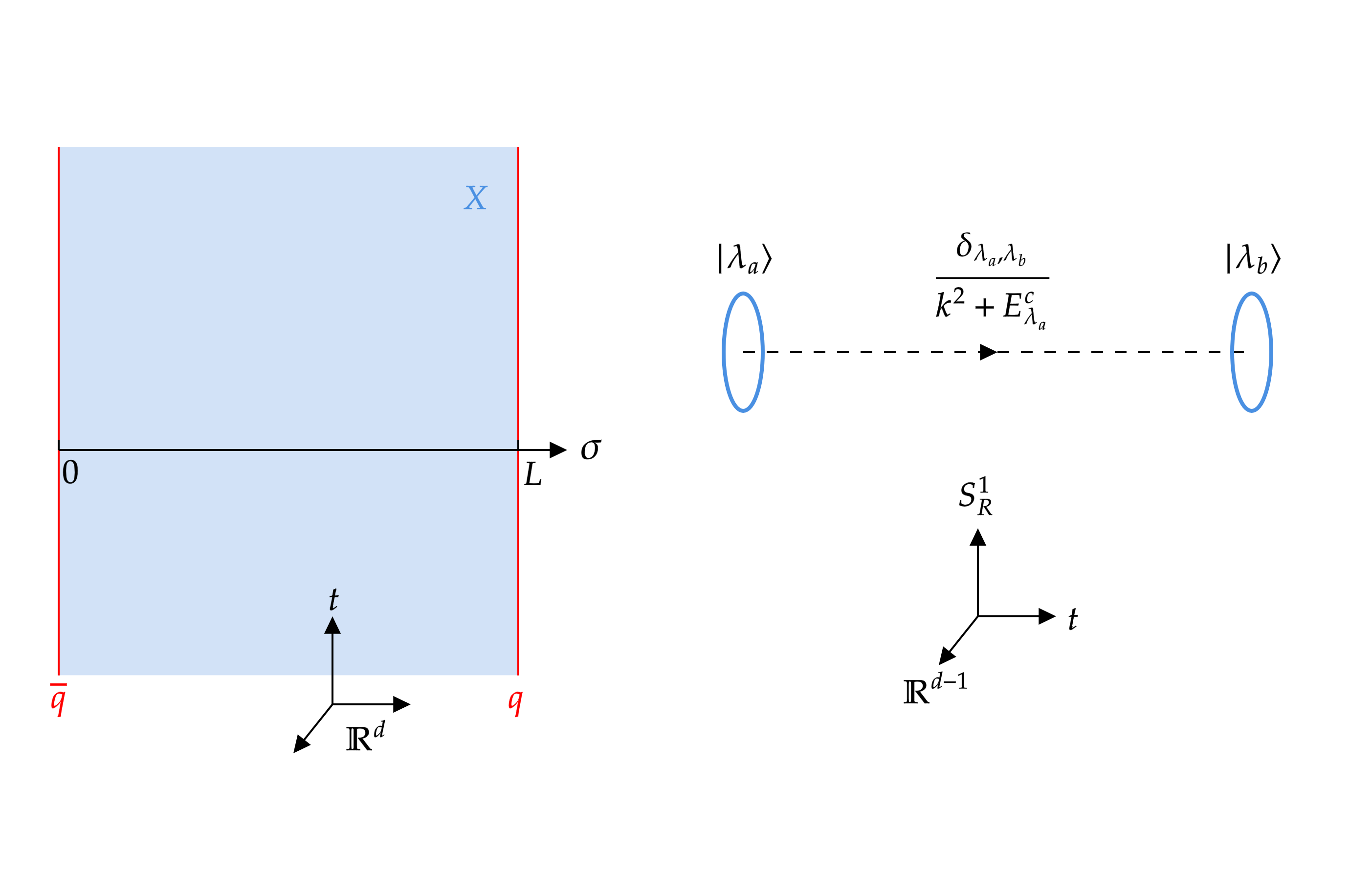}
  \caption{\label{pic_meson}The ``Meson" Configuration. In the open channel (Left), a single string is connected to Dirichlet boundaries (quark-antiquark pair) at $\sigma=0$ and $\sigma=L$. In the closed channel (Right), closed string states propagate in spacetime.}
\end{figure}

In the open channel, each time slice of the world sheet is an open string as in figure \ref{pic_meson}, and \eqref{eq_meson partition function} is interpreted as the thermal partition function 
\begin{equation}
\label{eq_meson open channel}
\begin{aligned}
\mathcal{Z}_{\text{m}}=\sum_{\lambda} e^{-2\pi R E_\lambda^{\text{o}}}
    \end{aligned}
\end{equation}
Indeed \eqref{eq_meson open channel} from the open string spectrum agrees with \eqref{eq_meson partition function}. On the other hand, we may perform a Wick rotation and take the time to be horizontal in figure~\ref{pic_meson}. In this case, \eqref{eq_meson partition function} is interpreted as the two point function of Polyakov loops $\Omega$ and $\Omega^*$, separated by $\vec{X}=(L,0,...,0)$. When $\Omega$ acts on the vacuum it creates a certain combination of closed string energy eigenstates, each of which behaves as a massive particle in $\mathbb{R}^d$: 
\begin{equation}
\label{eq_Polyakov Loop decomposition}
\begin{aligned}
|\Omega\rangle=&\sum_{\lambda} v_\lambda |\lambda\rangle\text{, such that }\\
\langle {\lambda}'(\vec{X}) | \lambda(0)\rangle=&\delta_{\lambda, {\lambda}'}\frac{(E^{\text{c}}_\lambda)^{\frac{d}{2}}l_s^{d-1}}{\sqrt{\pi}(2L)^{\frac{d-2}{2}}}K_{\frac{d-2}{2}}\left(E^{\text{c}}_\lambda L\right)
    \end{aligned}
\end{equation}
where we have used the massive propagator in $\mathbb{R}^d$ \cite{luscher2004string,aharony2011effective}.Therefore, the closed channel representation of \eqref{eq_meson partition function} reads \footnote{In the $SU(3)$ Yang-Mills theory, $\Omega$ should be understood as the trace of a Wilson line in the fundamental representation.}
\begin{equation}
\label{eq_meson closed channel rep}
\begin{aligned}
\mathcal{Z}_{\text{m}}&=\vev{\Omega^*({\vec{X}})\Omega(0)}\\
&=\sum_{\lambda}|v_\lambda|^2\frac{(E^{\text{c}}_\lambda)^{\frac{d}{2}}l_s^{d-1}}{\sqrt{\pi}(2L)^{\frac{d-2}{2}}}K_{\frac{d-2}{2}}\left(E^{\text{c}}_\lambda L\right)
    \end{aligned}
\end{equation}
Consistency requires that there exists a set of the $v_\lambda\in \mathbb{C}$ such that together with the closed string energies 
$E_\lambda^\text{c}$ we can match \eqref{eq_meson partition function} with \eqref{eq_meson closed channel rep}.  This is in general a highly nontrivial condition that constrains the effective action and relates the closed string spectrum with the open string spectrum. However to the order we are working, we can see that appropriate $v_\lambda$ exists by performing a modular transformation using the identity in \eqref{eq_modular properties}, 
\begin{equation}
\label{eq_meson partition function dual}
\begin{aligned}
\mathcal{Z}_{\text{m}}=&\frac{(\pi R/L)^{\frac{d-1}{2}}e^{-\mu L}}{\left[\eta(\Tilde{q})\right]^{(d-1)}}+O\left(1/R^2\right)
    \end{aligned}
\end{equation}
where $\Tilde{q}=e^{-\frac{2L}{R}}$ is the dual modular parameter. Note that the closed string states which can be created from $\Omega$ acting on vacuum have $n_{\lambda}^{\text{L}}=n_{\lambda}^{\text{R}}=n_{\lambda}$, as they do not carry longitudinal momentum. Additionally, they are singlets under the rotations in the transverse plane.
 We take $L, R\gg l_s$ and fixed ratio $L/R$.  The ratio $L/R$ could be large or small, convenient for series expansion in the closed or the open channel, respectively.
In this limit, the closed channel representation \eqref{eq_meson closed channel rep} admits the following expansion 
\begin{equation}
\label{eq_meson closed channel rep expand}
\begin{aligned}
\mathcal{Z}_{\text{m}}=&\frac{\left(\pi R/L\right)^{\frac{d-1}{2}}e^{-\mu L}}{\Tilde{q}^{\frac{d-1}{24}}}\sum_\lambda|v_\lambda|^2\Tilde{q}^{n_\lambda}+\left(1/R^2\right)
    \end{aligned}
\end{equation}
we can choose the convention such that $v_\lambda>0$, and by comparing \eqref{eq_meson partition function dual} with \eqref{eq_meson closed channel rep expand} we obtain 
\begin{equation}
\label{eq_Polyakov operator coefficients}
\begin{aligned}
v_\mathbf{0}=&1+O\left(1/R^2\right)\\
v_\mathbf{1}=&\sqrt{d-1}\left[1+O\left(1/R^2\right)\right]
    \end{aligned}
\end{equation}
etc., where the order 0 result is also known from the CFT literature \cite{ishibashi1989boundary} as the Dirichlet boundary state.

\section{The ``Baryon" in the Open Channel}
In this section, we study the three strings tied at a junction as in figure \ref{pic_setup}. The string endpoints in $\mathbb{R}^d$ are positioned at $\vec{X}=(L,0,...,0)$, $\vec{Y}=(\frac{L}{2},\frac{\sqrt{3}L}{2},...,0)$, and $\vec{Z}=(-\frac{L}{2},\frac{\sqrt{3}L}{2},...,0)$ with Dirichlet boundary conditions on the vertices of the equilateral triangle (see \cite{andreev2016some} for a discussion of the collinear case.). Classically, strings are straight lines meeting at the origin, which is the Fermat–Torricelli point that minimizes the sum of distances to the vertices. Unlike the EFT boundaries we reviewed above, the vertex's location oscillates along the longitudinal directions $X_1$, $Y_1$, and $Z_1$ of the strings. For instance, let $X_{1}=l_s x_1(t)$ be the fluctuating position of the vertex in the longitudinal direction of the first string, then the Nambu-Goto action \eqref{eq_Nambu-Goto Action} in static gauge (see appendix \ref{appendix_Gauge Fixing}) reads
\begin{multline}
\label{eq_gauge fixed}
S_X=\int dt \int_{l_s x_1(t)}^{L}d\sigma\left[-\frac{1}{l_s^2}+\frac{1}{2}(\partial_t x_i)^2\right.\\
\left.-\frac{1}{2}(\partial_\sigma x_i)^2+O\left(1/L^4\right)\right] 
\end{multline}
and similarly for the $Y$- and $Z$-worldsheets.
We see that to leading order one only has to modify the integration domain in the action by the time-dependent function $l_s x_1(t)$.

The vertex is point-like in the IR, and the rigid geometric condition at the vertex $\sigma=0$ is such that $x_1=(z_2-y_2)/\sqrt{3}$, $y_1=(x_2-z_2)/\sqrt{3}$, $z_1=(y_2-x_2)/\sqrt{3}$, and for the transverse fluctuations
\begin{equation}
\label{eq_rigid vertex condition}
\begin{aligned}
\begin{cases}
x_2+y_2+z_2=0\\
x_j=y_j=z_j \text{, for }3\leq j \leq d 
\end{cases}
    \end{aligned}
\end{equation}
Since the longitudinal fluctuations $x_1$, $y_1$, and $z_1$ are related to the transverse fluctuations, we can easily find the normal modes and perform a perturbative expansion.

The string bulk action includes an order $-2$ classical part $S_\text{b}^{(-2)}=-\frac{3L}{l_s^2}\int dt$, and an order 0 quadratic part 
\begin{equation}
\label{eq_baryon 0th order action}
\begin{aligned}
S_{\text{b}}^{(0)} =&\frac{1}{2}\int_{\mathbb{R}\times[0,L]} d t d \sigma \left[(\partial_t x_i)^2-(\partial_\sigma x_i)^2+\text{cyclic}\right]\\
=&\frac{1}{2}\sum_{a=1}^3\int_{\mathbb{R}\times[0,L]} d t d \sigma \left[(\partial_t \xi_i^{[a]})^2-(\partial_\sigma \xi^{[a]}_i)^2\right]
\end{aligned}
\end{equation}
where we have applied a field redefinition $\xi_i^{[1]}=(x_i+y_i+z_i)/\sqrt{3}$, $\xi_i^{[2]}=(x_i-y_i)/\sqrt{2}$, and $\xi_i^{[3]}=(x_i+y_i-2z_i)/\sqrt{6}$ to diagonalize the system. At order 0, \eqref{eq_rigid vertex condition} is the Neumann boundary condition for $\xi_2^{[2,3]}$, $\xi_{j\geq 3}^{[1]}$ and the Dirichlet boundary condition for $\xi_2^{[1]}$, $\xi_{j\geq 3}^{[2,3]}$. 
In summary, the junction in the equilateral case behaves as a tensor product of Neumann and Dirichlet boundaries to the leading order 
\begin{equation}
\label{tensorp} 
(\text{Neumann})^{\otimes d}\otimes(\text{Dirichlet})^{\otimes (2d-3)}.
\end{equation}
Note that when $d=3$, Neumann and Dirichlet conditions are assigned to an equal number of polarizations, which will have some consequences below. Physically, $\xi_2^{[2,3]}$ and $\xi_{j\geq 3}^{[1]}$ are the spatial displacement, while $\partial_\sigma\xi_2^{[1]}$ and $\partial_\sigma\xi_{j\geq 3}^{[2,3]}$ are small rotation angles of the vertex.

At the vertex, we can write down an order $-1$ constant action $S_{\text{b}}^{(-1)}=-M\int_{\sigma=0} dt$, where $M$ is interpreted as the vertex mass as in \eqref{eq_ground state energy}. The non-linear realization of the Lorentz group requires $S_{\text{b}}^{(-1)}$ to be accompanied by an order 1 quadratic term 
\begin{multline}
\label{eq_baryon 1st order action}
S_\text{b}^{(1)}=\frac{Ml_s^2}{3}\int_{\sigma=0} dt\left[(\partial_t\xi_2^{[2]})^2+(\partial_t\xi_2^{[3]})^2\right]\\
+\frac{Ml_s^2}{6}\int_{\sigma=0} dt(\partial_t\xi_j^{[1]})^2
\end{multline}
which is unique and agrees with the expansion of a standard world-line action. Note that expanding \eqref{eq_gauge fixed} with respect to the longitudinal fluctuation yields another order 1 term, which is cubic in the fluctuations
\begin{alignb}
\label{eq_baryon 1st order cubic}
\Tilde{S}_\text{b}^{(1)}=&-\frac{l_s}{2}\int_{\sigma=0} dt \left\{x_1\left[(\partial_t x_i)^2-(\partial_\sigma x_i)^2\right]+\text{cyclic}\right\} \\
=&-\frac{l_s}{2\sqrt{6}}\int_{\sigma=0} d t\left\{\xi_{2}^{[2]}\left[(\partial_t \xi_2^{[3]})^2-(\partial_t \xi_2^{[2]})^2\right.\right.\\
&\left.-(\partial_\sigma \xi_j^{[3]})^2+(\partial_\sigma \xi_j^{[2]})^2\right]+2\xi_{2}^{[3]}\left(\partial_t \xi_{2}^{[2]}\partial_t\xi_{2}^{[3]}\right.\\
&\left.\left.-\partial_\sigma \xi_{j}^{[2]}\partial_\sigma\xi_{j}^{[3]}\right)\right\}
\end{alignb}
\noindent We remark that \eqref{eq_baryon 1st order cubic} has important implications at higher orders, but here it will not play any further role. 

To obtain how~\eqref{tensorp} is modified due to the mass of the junction~\eqref{eq_baryon 1st order action}, we recompute the dispersion relation of the polarizations $\xi_2^{[2,3]}$, $\xi_{j\geq 3}^{[1]}$ to find
\begin{equation}
\label{eq_massive vertex quantization equation}
\begin{aligned}
\cos (\omega L)=c_{\parallel,\perp}\omega \sin(\omega L)
    \end{aligned}
\end{equation}
where $c_\parallel=\frac{2Ml_s^2}{3}$ for planar modes $\xi_{2}^{[2,3]}$, $c_\perp=\frac{Ml_s^2}{3}$ for vertical modes $\xi_{j}^{[1]}$, and $\omega$ is the frequency. Depending on the scale and sign of $M$, there are essential differences in solutions to \eqref{eq_massive vertex quantization equation}:
\begin{figure}[h!]
\centering
\includegraphics[width=0.5\textwidth ]{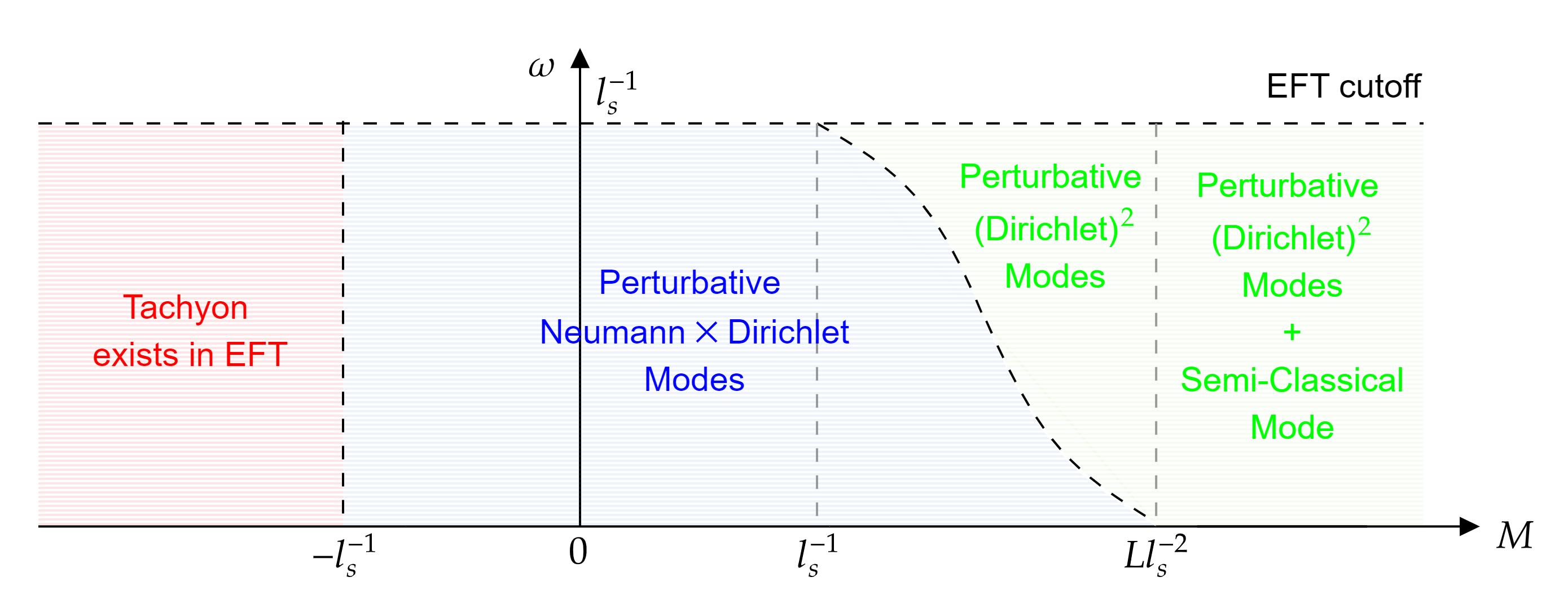}
  \caption{\label{pic_regimes of M open channel}The spectrum as a function of the vertex mass and system size.}
\end{figure}
\begin{itemize}
\item{$M<0$:} When $|M|\ll \frac{L}{l_s^2}$, \eqref{eq_massive vertex quantization equation} admits an imaginary solution $\omega\approx i|c_{\parallel,\perp}|^{-1}$. Such a tachyon is out of the EFT regime as long as $|M|\lesssim \frac{1}{l_s}$, where the stability needs to be examined in the full non-linear theory. However, if the junction mass is negative and parametrically large $|M|\gg \frac{1}{l_s}$, the instability is perturbative and within the EFT. We do not discuss the end-point of such instability, but we will get back to the question of whether slightly negative mass $-\frac{1}{l_s}\lesssim M<0 $ is indeed allowed. Interestingly, a negative baryon vertex mass is also found in certain large-$N$ gauge theories \cite{Imamura:2004tf}, with the property $|M|\lesssim \frac{1}{l_s}$ \footnote{The analogous configuration in large-$N$ theories is where three $\frac{N}{3}$-composite strings are connected at the junction. The composite string tension as well as the junction mass scale as $O(N)$ \cite{Witten:1979kh}, while the EFT cutoff depends on the fundamental string tension that scales as $O(N^0)$. In the large-$N$ limit, our discussion applies to composite string tension and junction mass in units of $N$, and the fact that the composite string tension does not set the cutoff scale indicates no perturbative instability.}. Anecdotally, a remote analogue is that soap films form plateau borders with a negative tension~\cite{teixeira2007energy} - of course, one need not worry about zero temperature instabilities in that setup.
\item{$M\sim \frac{1}{l_s}$:} In this case, solutions to \eqref{eq_massive vertex quantization equation} are Neumann-like with small corrections:
\begin{equation}
\label{eq_Neumann perturbative expansion}
\begin{aligned}
\quad \quad \omega_r=r\frac{\pi}{L}\left[1-\frac{c_{\parallel,\perp}}{L}+O\left(1/L^2\right)\right] 
    \end{aligned}
\end{equation}
where $r\in \mathbb{N}+\frac{1}{2}$. This is the most relevant regime to Yang-Mills theory, where we do not expect a hierarchy between the baryon junction mass and the string tension.
\item{$\frac{1}{l_s} \ll M \ll \frac{L}{l_s^2}$:} The spectrum of polarizations $\xi_2^{[2,3]}$ and $\xi_{j\geq 3}^{[1]}$ is divided into two regimes: low energy modes with $\omega \ll \frac{1}{Ml_s^2}$ follow the dispersion \eqref{eq_Neumann perturbative expansion}, while high energy modes with $\frac{1}{Ml_s^2} \ll \omega \ll \frac{1}{l_s} $ admit another expansion:
\begin{equation}
\label{eq_Dirichlet perturbative expansion}
\begin{aligned}
\quad \quad \omega_n=n\frac{\pi}{L}+\frac{1}{n\pi c_{\parallel,\perp}}+O\left(1/n^3\right)
    \end{aligned}
\end{equation}
where $n \in \mathbb{N} \gg \frac{L}{Ml_s^2}$. For those $\omega \gg \frac{1}{Ml_s^2}$ modes, the vertex condition become approximately Dirichlet for $\xi_2^{[2,3]}$ and $\xi_{j\geq 3}^{[1]}$.
\item{$M\gg \frac{L}{l_s^2}$:} In addition to the Dirichlet-Dirichlet modes as in \eqref{eq_Dirichlet perturbative expansion}, there exists low-frequency semi-classical modes with $\omega\approx (c_{\parallel,\perp}L)^{-1/2}$. These modes correspond to a heavy vertex oscillating in the classical potential without creating waves on the string. 
\end{itemize}
These regimes are summarized in figure \ref{pic_regimes of M open channel}. In the following, we will focus on the regime $M\sim l_s^{-1}$ where the string fluctuations dominate the physics. We expect $M\sim l_s^{-1}$ to be the case in Yang-Mills theory. Let us mention that $M\gg l_s^{-1}$ might be interesting as well for other applications; a physical example of a heavy junction could be in a variant of the Abelian Higgs model, where the one-form symmetry is broken to $\mathbb{Z}_3$ via a heavy charge 3 monopole.

\subsection{Partition Function}
Applying the method of subsection \ref{sec_meson}, we calculate the thermal partition function of the junction by compactifying time on $\mathbb{S}_R^{1}$. We first consider the classical action $S_{\text{b}}^{(-2)}=\frac{6\pi R L}{l_s^2}=3\mu L$, $S_{\text{b}}^{(-1)}=2\pi R M$, and the quadratic fluctuations \eqref{eq_baryon 0th order action}: 
\begin{equation}
\begin{aligned}
\mathcal{Z}_{\text{b}}^{(0)}=&e^{-S_{\text{b}}^{(-2)}-S_{\text{b}}^{(-1)}}\int \mathcal{D}x_i\mathcal{D}y_i\mathcal{D}z_i e^{-S_{\text{b}}^{(0)}}\\
=&\frac{e^{-3\mu L-2\pi R M}}{\left[\eta(\sqrt{q})\right]^{d}\left[\eta(q)\right]^{d-3}}
    \end{aligned}
\end{equation}
where $q=e^{-\frac{2\pi^2R}{L}}$. This is the thermal partition function to the order 0, where the junction can be treated by the tensor product of Neumann and Dirichlet boundaries~\eqref{tensorp}. 

In the long-string limit, the order 1 action can be treated perturbatively.
We have at this order a contribution from
\eqref{eq_baryon 1st order action} and \eqref{eq_baryon 1st order cubic}:
\begin{equation}
\label{eq_baryon partition function}
\begin{aligned}
\mathcal{Z}_{\text{b}}=\mathcal{Z}_{\text{b}}^{(0)}\left[1-\vev{S^{(1)}_\text{b}}-\vev{\Tilde{S}^{(1)}_\text{b}}+O\left(1/L^2\right)\right]
    \end{aligned}
\end{equation}
where $\vev{...}$ is the vacuum expectation value in the order 0 theory. Note that the cubic term $\vev{\Tilde{S}^{(1)}_\text{b}}=0$ as it is odd under parity. Furthermore, we remark that the cubic operator $\Tilde{S}^{(1)}_\text{b}$ does not perturb the open channel spectrum at order 1. 

To work out $\vev{S^{(1)}_\text{b}}$, We denote by $\Tilde{G}(\Sigma, {\Sigma}')$ the free field worldsheet propagator with Dirichlet condition at $\sigma=0$ and Nuemann condition at $\sigma=L$, and $\Tilde{G}_{\alpha \beta}\equiv \lim_{\Sigma\to {\Sigma}'} \partial_{\Sigma_\alpha}\partial_{{\Sigma}'_\beta}\Tilde{G}(\Sigma, {\Sigma}')$ its coincident point function. By the Wick theorem, we obtain the 1-loop result
\begin{equation}
\label{eq_baryon partition function piece 1}
\begin{aligned}
\vev{ S_\text{b}^{(1)}}=&\frac{(d+2)Ml_s^2}{6}\int_{\sigma=0} d\tau \Tilde{G}_{\tau \tau}\\
=&\frac{(d+2)Ml_s^2}{144L}\log q \left[2E_2(q)-E_2(\sqrt{q})\right]
    \end{aligned}
\end{equation}
where $E_2(q)$ is the Eisenstein series, see also appendix \ref{appendix_Regularization}. As a consistency check, \eqref{eq_baryon partition function piece 1} agrees with the ground state energy \eqref{eq_ground state energy} and perturbed frequencies \eqref{eq_Neumann perturbative expansion}. 

For what comes next, we remark that the open channel partition function also admits a dual representation
\begin{equation}
\label{eq_open channel dual}
\begin{aligned}
\mathcal{Z}_{\text{b}}^{(0)}=&\frac{(\pi R/L)^{d-\frac{3}{2}}e^{-3\mu L-2\pi R M}}{2^{d/2}\left[\eta(\Tilde{q}^2)\right]^{d}\left[\eta(\Tilde{q})\right]^{d-3}}\\
\vev{ S_\text{b}^{(1)}}=&-\frac{(d+2)Ml_s^2}{36R}\left[2E_2(\Tilde{q}^2)-E_2(\Tilde{q})\right]
    \end{aligned}
\end{equation}
following from \eqref{eq_modular properties} and $\tilde{q}=e^{-\frac{2L}{R}}$.

\section{The String Interaction Vertex}

In this section, we discuss the closed channel interpretation of \eqref{eq_open channel dual} as a correlation function of Polyakov loops $\Omega$ in $\mathbb{R}^{d}\times\mathbb{S}^1_R$. From the point of view of the non-compact $\mathbb{R}^d$, the Polyakov loops are point operators and the 1-form $\mathbb{Z}_3$ is reduced to a 0-form $\mathbb{Z}_3$, with the fundamental loop carrying charge $1\ {\rm mod} \ 3$. Symmetry preserving three-point functions are $\langle\Omega \Omega \Omega\rangle$ and $\langle\Omega^* \Omega^* \Omega^*\rangle$. We will now explain how to reinterpret the junction partition function in terms of the three-point function 
\begin{equation}
\label{eq_Polyakov three-point function}
\begin{aligned}
\mathcal{Z}_{\text{b}}=\langle \Omega(\vec{X})\Omega(\vec{Y})\Omega(\vec{Z})\rangle
    \end{aligned}
\end{equation}
If we take the time direction to be along $\mathbb{R}^d$ then $\langle \Omega(\vec{X})\Omega(\vec{Y})\Omega(\vec{Z})\rangle$ describes an interaction vertex of three closed strings. As in \eqref{eq_Polyakov Loop decomposition}, we start from the idea that the $\Omega$ operator creates a superposition of energy eigenstates when acting on the vacuum. The coefficients of the energy eigenstates are \eqref{eq_Polyakov operator coefficients}. 
When $R,L\gg l_s$, the closed string states are heavy particles which travel for a long distance. We assume therefore that in the limit $L, R\gg l_s$, \eqref{eq_Polyakov three-point function} can be interpreted as strings scattering via a local interaction. We will further show that to the order we compute, such interaction is purely a contact interaction and the interaction strength $C_{\lambda_{x}\lambda_{y}\lambda{z}}$ between the energy eigenstates $\lambda_{x}$, $\lambda_{y}$, and $\lambda_{z}$ can be determined unambiguously.

We remark that the non-locality of the scattering, which we expect to be on the scale $l_s$ modulo logarithmic corrections,  leads to higher contact couplings between the states $\lambda$'s. From Lorentz invariance, it follows that these non-s-wave scatterings contribute at least at order 2 and hence can be neglected for the analysis here. We conclude that we expect the partition function to be reproduced by a simple contact interaction between the string states: 
\begin{equation}
\begin{aligned}
\mathcal{Z}_{\text{b}}=\mathcal{Z}_{\text{s-wave}}+O\left(1/R^2\right)
    \end{aligned}
\end{equation}
Let $\vec{W} \in \mathbb{R}^d$ be the location of the contact interaction vertex and $L_{xw}=|\vec{X}-\vec{W}|$, the s-wave amplitude is an integration against the propagators \eqref{eq_Polyakov Loop decomposition} that reads
\begin{multline}
\label{eq_s-wave integral}
\mathcal{Z}_{\text{s-wave}}=\sum_{\lambda_{x,y,z}}\int \frac{d^d\vec{W}}{l_s^d}C_{\lambda_x\lambda_y\lambda_z} \times \\
\prod_{a=x,y,z}\left[\frac{v_{\lambda_a}(E^{\text{c}}_{\lambda_a})^{\frac{d}{2}}l_s^{d-1}}{\sqrt{\pi}(2L_{wa})^{\frac{d-2}{2}}}K_{\frac{d-2}{2}}\left(E^{\text{c}}_{\lambda_a} L_{wa}\right)\right]~.
\end{multline}
This is just the tree-level diagram in a theory in $\mathbb{R}^d$ with cubic interactions 
\begin{equation}
\label{intvert}
\sim\sum_{\lambda_{x,y,z}}\int \frac{d^d\vec{W}}{l_s^d}C_{\lambda_x\lambda_y\lambda_z}\Phi_{\lambda_x}\Phi_{\lambda_y}\Phi_{\lambda_z}~.
\end{equation} 
The fields $\Phi_{\lambda_a}$ are the string fields in $\mathbb{R}^d$ that create the energy eigenstates $|\lambda_a\rangle$ wrapped on $\mathbb{S}^1_R$. These fields have mass  $E^{\text{c}}_{\lambda_a}$  and hence the propagators as in \eqref{eq_Polyakov Loop decomposition} and \eqref{eq_s-wave integral}. 

The integral~\eqref{eq_s-wave integral} is heavily dominated by a saddle point near the origin, similar to \cite{dymarsky20183d, Dymarsky:2013wla}. To the order we are concerned, the saddle point solves a generalization of the Fermat-Torricelli problem where each edge is weighted by $E_{\lambda_{a}}^{\text{c}}$ given in \eqref{eq_closed string energy}. The saddle point value reads
\begin{multline}
F_{\lambda_{x}\lambda_{y}\lambda_{z}}=\max_{\Vec{W}}\left[\prod_{a}\frac{(E^{\text{c}}_{\lambda_a})^{\frac{d}{2}}l_s^{d-1}}{\sqrt{\pi}(2L_{wa})^{\frac{d-2}{2}}}K_{\frac{d-2}{2}}\left(E^{\text{c}}_{\lambda_a} L_{wa}\right)\right]\\
=\left(\frac{\pi R}{L}\right)^{\frac{3(d-1)}{2}}\frac{e^{-3\mu L}}{\Tilde{q}^{\frac{d-1}{8}}}\Tilde{q}^{n_x+n_y+n_z}\left[1+O\left(1/R^2\right)\right]
\end{multline}
Around the saddle point, we take the Gaussian integral -- this is crucial to the order we compute, however, corrections to the Gaussian integral are of order 2 and higher. 

We find that the s-wave amplitude agrees with the open channel partition function in its leading $L$-dependence:
\begin{multline}
\label{eq_s-wave result}
\mathcal{Z}_{\text{s-wave}}=(\pi R/L)^{d-\frac{3}{2}}\frac{2\pi^d e^{-3\mu L}}{3^{\frac{d}{2}}\Tilde{q}^{\frac{d-1}{8}}}\times\\
\sum_{\lambda_{x,y,z}}C_{\lambda_x\lambda_y\lambda_z}v_{\lambda_x} v_{\lambda_y}v_{\lambda_z}\Tilde{q}^{n_x+n_y+n_z}\left[1+ O \left(1/R^2\right) \right]
\end{multline}
By comparing \eqref{eq_s-wave result} with \eqref{eq_baryon partition function} and \eqref{eq_open channel dual}, we found the cubic coupling between the three lowest-lying string states $\mathbf{0}$: 
\begin{equation}
\label{eq_s-wave coupling 1}
\begin{aligned}
C_{\mathbf{0}\mathbf{0}\mathbf{0}}=\frac{e^{-2\pi M R}}{2(2\pi^2/3)^{\frac{d}{2}}}\left[1+\frac{(d+2) M l_s^2}{36 R}+O\left(1/R^2\right)\right]
    \end{aligned}
\end{equation}
We can also identify the coupling between the two lowest-lying ones and the second-lowest string $\mathbf{1}$:
\begin{multline}
\label{eq_s-wave coupling 2}
C_{\mathbf{0}\mathbf{0}\mathbf{1}}=\frac{e^{-2\pi M R}}{2(2\pi^2/3)^{\frac{d}{2}}}\left[\frac{d-3}{3\sqrt{d-1}}\right. \\
\left.+\frac{(d+2) (d+21) M l_s^2}{108 \sqrt{d-1} R}+O\left(1/R^2\right)\right]
\end{multline}
At the order we are computing, among the higher string states there are degeneracies and one cannot distil the interaction vertices of each state. Instead, one can obtain average predictions from \eqref{eq_open channel dual} and \eqref{eq_s-wave integral}. At the next order which we do not elaborate on here, it is possible to go further.

\begin{figure}[h!]
\centering
\includegraphics[width=0.4\textwidth ]{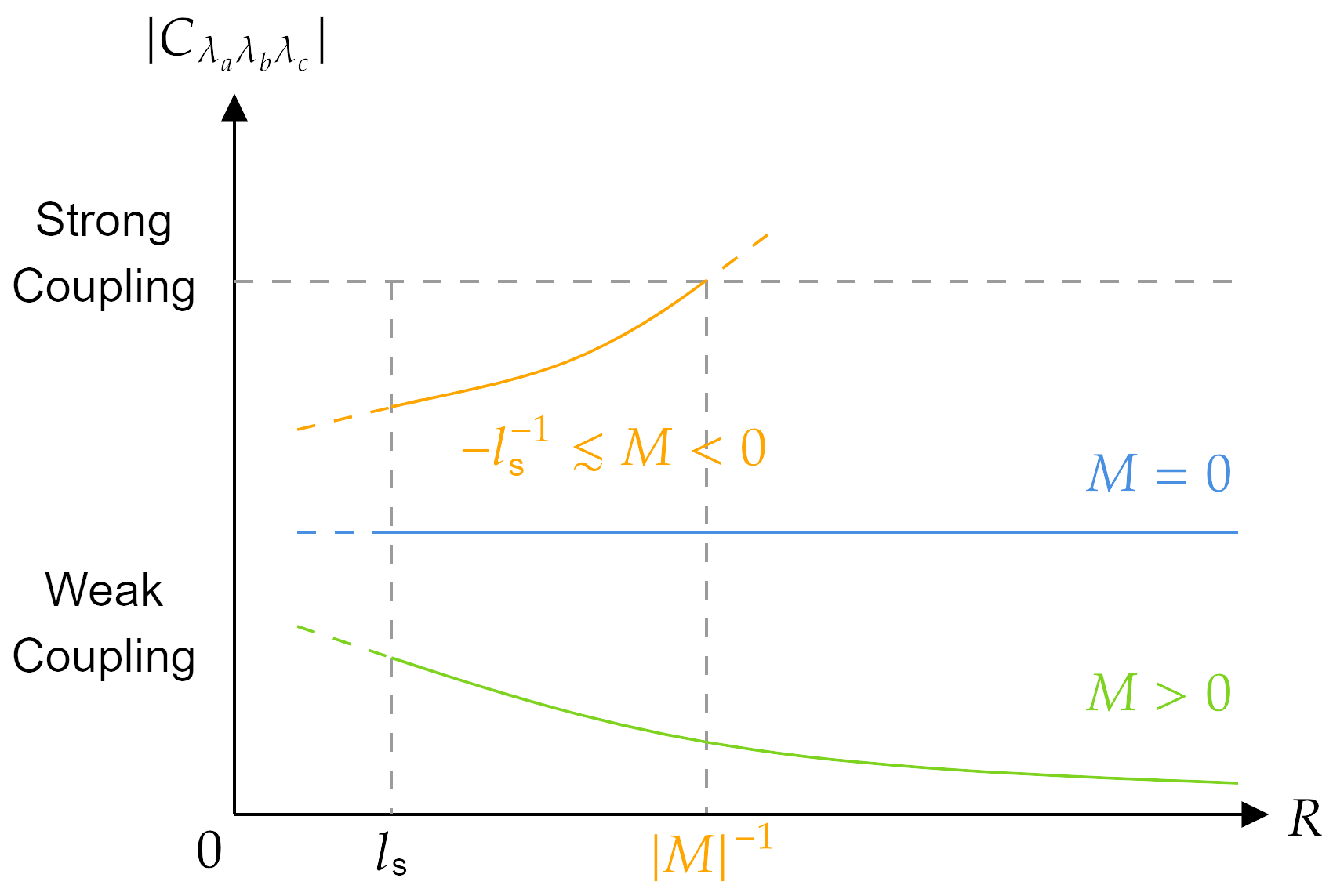}
  \caption{The contact interaction with various $M$ and $R$. The EFT breaks down when $R\lesssim l_s$ (left dashed lines), and the perturbative description fails for long strings if $M<0$ (upper dashed line).}
\end{figure}

One immediate lesson is that the overall strength of the interaction among closed strings is \begin{equation} C_{\lambda_{a}\lambda_{b}\lambda_{c}}\sim e^{-2\pi M R}~.\end{equation} 
This is essentially the string coupling constant among strings of length $2\pi R$. It can be interpreted that the cubic interaction among segments of size $l_s$ is $e^{-2\pi M l_s}$, and such a probability is raised to the power of the number of segments, $2\pi R/l_s$. These estimates should be general and applicable to other cases, for instance, excited glueballs \footnote{For critical strings, the cubic couplings were recently discussed in~\cite{rosenhaus2022chaos}.}. We remark that in QCD where strings (flux tubes) are breakable, the EFT predictions are still reliable as far as the string is concerned~\cite{Bonati:2021vbc}.

For $M>0$, we see that the interaction is extremely weak for long closed strings while for $-\frac{1}{l_s}\lesssim M<0$, the interaction becomes strong when $R\gg\frac{1}{|M|}$. We comment that a negative $M$ that is parametrically larger than $l_s^{-1}$ is ruled out by perturbative stability and $-\frac{1}{l_s}\lesssim M<0$ leads to strong couplings between long strings and hence requires further analysis of unitarity.

A central assumption is that when we act with the Polyakov loop $\Omega$ on the vacuum we create single-string eigenstates. This assumption appears to be jeopardized when the cubic coupling is strong. As far as we are aware, lattice simulations, including the extensive simulations we referred to in the introduction, showed no sign that the Polyakov loop mixes single string and two (anti) string states. Therefore, it follows that $-\frac{1}{l_s}\lesssim M<0$ is strongly disfavored by unitarity arguments and one is compelled to suggest $M\geq 0$. It remains puzzling that in some large $N$ gauge theories negative values of $M$ were reported in~\cite{Imamura:2004tf}.

When $d=3$ and $M=0$, we notice that \eqref{eq_open channel dual} contains only even power of $\Tilde{q}$. This is a consequence of a chiral $\mathbb{Z}_2$ symmetry of the NGBs. Explicitly in the open channel, this chiral $\mathbb{Z}_2$ is a combination of the transformation $\partial_t x_i \to i\partial_\sigma x_i$, $\partial_\sigma x_i \to -i\partial_t x_i$, which is T-duality-like, and the spatial reflection $x_2\to x_3$, $x_3\to x_2$. Such a chiral $\mathbb{Z}_2$ symmetry is preserved by the string bulk action \eqref{eq_Nambu-Goto Expansion} on X-, Y-, and Z-worldsheets up to order 2, and at higher orders it is possible to write an EFT term that violates it. Note that the chiral $\mathbb{Z}_2$ exchanges the Neumann (Dirichlet) boundary condition in the $X_2$ direction with that of Dirichlet (Neumann) in the $X_3$ direction. Therefore, from \eqref{tensorp} we find this chiral $\mathbb{Z}_2$ is preserved by the vertex only when $d=3$.

In the closed channel, the Polyakov operators $\Omega$ create $\mathbb{R}^3$ scalars that are charged under the chiral $\mathbb{Z}_2$. Let $\alpha_{-n}^i$ ($\Tilde{\alpha}_{-n}^i$) be the closed string left (right) moving modes creation operators at order 0, where $n\in \mathbb{N}^+$ and $i=2,3$. A generic scalar state is created by acting with $(\alpha_{-n}\cdot \alpha_{-{n}'})$, $(\alpha_{-n}\cdot\Tilde{\alpha}_{-{n}'})$, and  $(\Tilde{\alpha}_{-n}\cdot\Tilde{\alpha}_{-{n}'})$ on the ground state. Obviously, a scalar state is not charged under the reflection, and the number of $\alpha_{-n}^i$ and $\Tilde{\alpha}_{-n}^i$ operators satisfy $m=\Tilde{m} \mod{2}$. The transformation  $\partial_t x_i \to i\partial_\sigma x_i$, $\partial_\sigma x_i \to -i\partial_t x_i$ when acting on mode operators reads
\begin{equation}
\begin{aligned}
\begin{cases}
\alpha_{-n}^i \to -\alpha_{-n}^i\\
\Tilde{\alpha}_{-n}^i \to \Tilde{\alpha}_{-n}^i
\end{cases}
    \end{aligned}
\end{equation}
Therefore, each scalar state under the chiral $\mathbb{Z}_2$ acquires $(-1)^m=(-1)^{\Tilde{m}}$ that depends on the number of left (right) moving mode operators. We conclude the s-wave coupling $C_{\lambda_{a}\lambda_{b}\lambda_{c}}$ vanishes up to order 2 if 
\begin{equation}
\begin{aligned}
m_{\lambda_a}+m_{\lambda_b}+m_{\lambda_c}=1\mod{2}
    \end{aligned}
\end{equation}
This has explained the selection rule we found when $d=3$ and $M=0$, where $m_\mathbf{0}=0$ and $m_\mathbf{1}=1$. We remark that this suggests $M$ is potentially a symmetry-breaking parameter in the $(3+1)$ dimension.

\section{Conclusion and Outlook}

In this paper, we have demonstrated the open-closed duality for a ``baryon'' configuration as in figure \ref{pic_setup}. We claim that up to order 1, the effective theory has two parameters: the string tension $l_s^{-2}$ and the vertex mass $M$. We specified the action and boundary conditions as in \eqref{eq_rigid vertex condition}, \eqref{eq_baryon 0th order action}, \eqref{eq_baryon 1st order action}, and \eqref{eq_baryon 1st order cubic}. 

We showed that in the closed channel the ``baryon'' is mapped to s-wave scattering of closed strings and we extracted the universal s-wave couplings \eqref{eq_s-wave coupling 1} and \eqref{eq_s-wave coupling 2}. These couplings have important implications for IR physics: We found that $M<0$ suggests strong coupling and possible unitarity violation, while $M>0$ implies weak coupling and is stable. Intriguingly, in (3+1) dimension the interaction is subject to a selection rule, which we argue is a consequence of a worldsheet chiral $\mathbb{Z}_2$ symmetry. We pointed out that $M\neq 0$ in (3+1) dimension breaks this symmetry and violates the selection rule.

Finally, we list a few important questions:

\begin{itemize}
\item{Non-equilateral configurations}: The interaction vertex~\eqref{intvert} is independent of the location from which the closed strings are propagating. This is a trivial consequence of locality. To test this simple fact we must
consider the case where the end-point quarks of the baryon junction are positioned on a non-equilateral triangle, i.e. such that $\vec{X}=(L_x,0,0...,0)$, $\vec{Y}=(-\frac{L_y}{2},\frac{\sqrt{3}L_y}{2},0...,0)$, and $\vec{Z}=(-\frac{L_z}{2},-\frac{\sqrt{3}L_z}{2},0...,0)$. At order 0 and when $M=0$, the quantization condition follows from \eqref{eq_baryon 0th order action} and the rigid condition \eqref{eq_rigid vertex condition}. For a planar mode, it reads 
\begin{equation}
\label{eq_non-equilateal piece 1}
\quad \quad \cos(\omega L_x)\sin(\omega L_y)\sin(\omega L_z)+\text{cyclic}=0~,
\end{equation}
while for a vertical mode:
\begin{equation}
\label{eq_non-equilateal piece 2}
\quad \quad \sin(\omega L_x)\cos(\omega L_y)\cos(\omega L_z)+\text{cyclic}=0~.
\end{equation}
In the open channel interpretation (i.e. the baryon junction channel), the partition function is evaluated as
\begin{equation}
\label{eq_non-equilateal piece 3}
\begin{aligned}
\mathcal{Z}_{\text{b}}=\sum_{\omega}e^{-2\pi R \omega }
    \end{aligned}
\end{equation}
which is a function of modular parameters $q_{x,y,z}=e^{-\frac{2\pi^2 R}{L_{x,y,z}}}$. On the other hand, in the closed channel, all we have to do in~\eqref{eq_s-wave integral} is to change the points from which we propagate the closed strings. The amplitudes $v_{\lambda_a}$ and $C_{\lambda_a\lambda_b\lambda_c}$ are insensitive to where the strings are coming from, by locality. Hence the closed channel prediction is
\begin{multline}
\label{eq_non-equilateral s-wave}
\quad \quad\mathcal{Z}_{\text{s-wave}}=\left(\frac{3\pi R}{L_x+L_y+L_z}\right)^{d-\frac{3}{2}}
\frac{2\pi^d e^{-3\mu L}}{3^{\frac{d}{2}}(\Tilde{q}_x\Tilde{q}_y\Tilde{q}_z)^{\frac{d-1}{24}}}
\\
\times \left[\frac{(L_x+L_y+L_z)^2}{3(L_xL_y+L_yL_z+L_zL_x)}\right]^{\frac{d}{2}-1}\\
\times\sum_{\lambda_{x,y,z}}C_{\lambda_x\lambda_y\lambda_z}v_{\lambda_x} v_{\lambda_y}v_{\lambda_z}\Tilde{q}_x^{n_{\lambda_x}}\Tilde{q}_y^{n_{\lambda_y}}\Tilde{q}_z^{n_{\lambda_z}}
\end{multline}
as a function of the dual variables $\Tilde{q}_{x,y,z}=e^{-\frac{2L_{x,y,z}}{R}}$.  We expect \eqref{eq_non-equilateral s-wave} to be identified with \eqref{eq_non-equilateal piece 3} following \eqref{eq_non-equilateal piece 1} and \eqref{eq_non-equilateal piece 2} through a multi-variable modular transformation. It would be nice to carry this out. On the other hand, if the $\Vec{X}\Vec{Y}\Vec{Z}$-triangle has an inner angle greater than $120^{\circ}$ then the Fermat point coincides with the obtuse vertex as in \cite{andreev2016some}, and a separate discussion of the junction condition and operators is necessary.

\item{Chiral $\mathbb{Z}_2$ symmetry}: It would be interesting to understand how exactly this symmetry is broken if $M=0$.
In \cite{Imamura:2004tf} it is suggested that $M=0$ in the large-$N$ Maldacena-Nunez solution. It would be nice to understand if the chiral $\mathbb{Z}_2$ symmetry we discussed is always an IR accidental symmetry or if it could be related to some microscopic symmetry. 
\item{Instability}: We encountered a perturbative instability when $M\lesssim -l_s^{-1}$. From an RG consideration, the endpoint of the instability cannot be a point-like junction. It would be interesting to find if there is a fat junction solution to the full nonlinear Nambu-Goto theory.

\item{Higher orders and non-s-wave scattering}: It is straightforward to push the precision of this paper to order 2 and higher. Up to (and including) order 2, $l_s$ and $M$ are the only two parameters of the EFT. We would like to know if at this order non-s-wave contact interactions appear.

\item It would be nice to know the junction mass in Yang-Mills theories and other similar theories and to test the theory we have discussed.
\end{itemize}

\begin{acknowledgments}
We thank O. Aharony, Gabriel Cuomo, David Frenklakh, Alberto Nicolis, Alessandro Podo and Amit Sever for useful comments.  ZK and SZ are supported in part by the Simons Foundation grant 488657 (Simons Collaboration on the Non-Perturbative Bootstrap), the BSF grant no. 2018204 and NSF award number 2310283.

\end{acknowledgments}

\appendix

\section{Gauge Fixing}
\label{appendix_Gauge Fixing}
We briefly explain the gauge we used in \eqref{eq_gauge fixed}, as it is not completely standard. First, we notice that  diffeomorphism invariance allows us to choose 
\begin{equation}
\begin{aligned}
\begin{cases}
    X_0=\Tilde{t}\\
X_1=\tilde{\sigma}+l_s f(\tilde{\sigma}) x_1(\Tilde{t})\\
X_i=l_sx_i(\Tilde{t},\tilde{\sigma})\text{, for }2\leq i\leq d
\end{cases}
    \end{aligned}
\end{equation}
where $\Tilde{\sigma}\in [0,L]$, $x_1(\tilde{t})$ is the longitudinal displacement, and $f(\Tilde{\sigma})$ is a smooth monotonic function such that $f(0)=1$ and $f(L)=0$.

As in the static gauge, we would like to pick a new coordinate $t=X_0=\tilde{t}$ and $\sigma=X_1=\tilde{\sigma}+l_s f(\tilde{\sigma}) x_1(\Tilde{t})$. Following the chain rule 
\begin{equation}
\begin{aligned}
\partial_t x_i=&\partial_{\tilde{t}}x_i-l_s f \partial_t x_1 \partial_{\Tilde{\sigma}}x_i\\
\partial_\sigma x_i=& (1+l_s\partial_{\tilde{\sigma}}f x_1)^{-1}\partial_{\tilde{\sigma}}x_i
    \end{aligned}
\end{equation}
one can verify that the bulk action is as in \eqref{eq_Nambu-Goto Expansion}, yet the domain of integration becomes dynamical $\sigma\in [l_s x(t),L]$.
\section{Green's Function and Regularization}
\label{appendix_Regularization}
The Dedekind $\eta$ function is defined as 

\begin{equation}
\begin{aligned}
\eta(q)\equiv q^{\frac{1}{24}}\prod_{n\in\mathbb{N}^+}(1-q^n)
    \end{aligned}
\end{equation}
The Eisenstein series is defined as 
\begin{equation}
\begin{aligned}
E_{2k}(q)\equiv 1+\frac{2}{\zeta (1-2k)}\sum_{n\in \mathbb{N}^+}\frac{n^{2k-1}q^n}{1-q^n}\text{, }k\in\mathbb{N}^+
    \end{aligned}
\end{equation}
Let $q=e^{2\pi i \tau}$ and $\Tilde{\tau}=-\frac{1}{\tau}$, we will use the following modular transformation of these functions 
\begin{equation}
\label{eq_modular properties}
\begin{aligned}
\eta(q)=&\sqrt{-i\Tilde{\tau}}\eta(\Tilde{q})\\
E_2(q)=&-\frac{6i}{\pi }\Tilde{\tau}+\Tilde{\tau}^2E_2(\Tilde{q})\\
E_{2k}(q)=&\Tilde{\tau}^{2k}E_{2k}(\Tilde{q})
    \end{aligned}
\end{equation}
More practically, we will need
\begin{equation}
\begin{aligned}
\log q E_2(q)=&-12-\log \Tilde{q} E_2(\Tilde{q})\\
\log q E_2(\sqrt{q})=&-24-4\log \Tilde{q} E_2(\Tilde{q}^2)
    \end{aligned}
\end{equation}

We denote the modular parameter as $q=e^{-\frac{2\pi^2 R}{L}}$, and a useful infinite sum reads
\begin{equation}
\label{eq_appendix_sum 1}
\begin{aligned}
\sum_{r\in \mathbb{N}+\frac{1}{2}}\sum_{m \in \mathbb {Z}}\frac{\frac{\pi^2r^2}{L^2}}{\frac{\pi^2r^2}{L^2}+\frac{m^2}{R^2}}=&-\frac{\log q}{2}\sum_{r\in \mathbb{N}+\frac{1}{2}}r\frac{1+q^r}{1-q^r}\\
=&-\frac{\log q}{48}\left[2E_2(q)-E_2(\sqrt{q})\right]\\
\sum_{r\in \mathbb{N}+\frac{1}{2}}\sum_{m \in \mathbb {Z}}\frac{\frac{m^2}{R^2}}{\frac{\pi^2r^2}{L^2}+\frac{m^2}{R^2}}=&-\sum_{r\in \mathbb{N}+\frac{1}{2}}\sum_{m \in \mathbb {Z}}\frac{\frac{\pi^2r^2}{L^2}}{\frac{\pi^2r^2}{L^2}+\frac{m^2}{R^2}}\\=&\frac{\log q}{48}\left[2E_2(q)-E_2(\sqrt{q})\right]
    \end{aligned}
\end{equation}
where we have used $\sum_{m\in \mathbb {Z}}1=1+2\zeta(0)=0$.

For the Neumann-Dirichlet boundary condition on $\mathbb{S}_R^1\times [0,L]$, the Green's function reads 
\begin{multline}
\label{eq_appendix_ND function}
\Tilde{G}(\sigma,{\sigma}',\tau-{\tau}')=\frac{1}{\pi R L}\sum_{r\in \mathbb{N}+\frac{1}{2}}\sum_{m \in \mathbb {Z}}\frac{e^{i\frac{m}{R}(\tau-{\tau}')}}{\frac{\pi^2r^2}{L^2}+\frac{m^2}{R^2}}\\
\times \cos\left(\frac{r\pi\sigma}{L}\right)\cos\left(\frac{r\pi{\sigma}'}{L}\right)
\end{multline}
The coincident point derivatives of \eqref{eq_appendix_ND function} are
\begin{equation}
\begin{aligned}
\Tilde{G}_{\tau \tau}(\sigma)=&\frac{1}{\pi R L}\sum_{r\in \mathbb{N}+\frac{1}{2}}\sum_{m \in \mathbb {Z}}\frac{\frac{m^2}{R^2}}{\frac{\pi^2r^2}{L^2}+\frac{m^2}{R^2}}\cos^2\left(\frac{r\pi\sigma}{L}\right)\\
    \end{aligned}
\end{equation}
Using \eqref{eq_appendix_sum 1}, we obtain
\begin{equation}
\label{eq_appendix_overlaps 3}
\begin{aligned}
\int_{\sigma=0} d\tau \Tilde{G}_{\tau \tau}&=\frac{2}{L}\sum_{r\in \mathbb{N}+\frac{1}{2}}\sum_{m\in\mathbb{Z}}\frac{\frac{m^2}{R^2}}{\frac{\pi^2r^2}{L^2}+\frac{m^2}{R^2}}\\
&=\frac{\log q}{24L}\left[2E_2(q)-E_2(\sqrt{q})\right]
    \end{aligned}
\end{equation}

\onecolumngrid
\quad
\twocolumngrid

\bibliographystyle{unsrt}
\bibliography{Ref}

\begin{thebibliography}{10}

\bibitem{Abrikosov1957zh}
AA~Abrikosov.
\newblock Zh. eksp. teor. fiz.
\newblock {\em Sov. Phys. JETP 5 1174}, 32:1442, 1957.

\bibitem{Nielsen:1973cs}
Holger~Bech Nielsen and P.~Olesen.
\newblock {Vortex Line Models for Dual Strings}.
\newblock {\em Nucl. Phys. B}, 61:45--61, 1973.

\bibitem{Polchinski:1991ax}
Joseph Polchinski and Andrew Strominger.
\newblock {Effective string theory}.
\newblock {\em Phys. Rev. Lett.}, 67:1681--1684, 1991.

\bibitem{luscher2004string}
Martin L{\"u}scher and Peter Weisz.
\newblock String excitation energies in su (n) gauge theories beyond the free-string approximation.
\newblock {\em Journal of High Energy Physics}, 2004(07):014, 2004.

\bibitem{Drummond:2004yp}
J.~M. Drummond.
\newblock {Universal subleading spectrum of effective string theory}.
\newblock 11 2004.

\bibitem{aharony2009effective}
Ofer Aharony and Eyal Karzbrun.
\newblock On the effective action of confining strings.
\newblock {\em Journal of High Energy Physics}, 2009(06):012, 2009.

\bibitem{aharony2010corrections}
Ofer Aharony and Nizan Klinghoffer.
\newblock Corrections to nambu-goto energy levels from the effective string action.
\newblock {\em Journal of High Energy Physics}, 2010(12):1--18, 2010.

\bibitem{aharony2011effective}
Ofer Aharony and Matan Field.
\newblock On the effective theory of long open strings.
\newblock {\em Journal of High Energy Physics}, 2011(1):1--48, 2011.

\bibitem{aharony2012effective}
Ofer Aharony and Matthew Dodelson.
\newblock Effective string theory and nonlinear lorentz invariance.
\newblock {\em Journal of High Energy Physics}, 2012(2):1--13, 2012.

\bibitem{dubovsky2012effective}
Sergei Dubovsky, Raphael Flauger, and Victor Gorbenko.
\newblock Effective string theory revisited.
\newblock {\em Journal of High Energy Physics}, 2012(9):1--21, 2012.

\bibitem{Meineri:2012uav}
Marco Meineri.
\newblock {Lorentz completion of effective string action}.
\newblock {\em PoS}, ConfinementX:041, 2012.

\bibitem{aharony2013effective}
Ofer Aharony and Zohar Komargodski.
\newblock The effective theory of long strings.
\newblock {\em Journal of High Energy Physics}, 2013(5):1--26, 2013.

\bibitem{Brambilla:2014eaa}
Nora Brambilla, Michael Groher, Hector~E. Martinez, and Antonio Vairo.
\newblock {Effective string theory and the long-range relativistic corrections to the quark-antiquark potential}.
\newblock {\em Phys. Rev. D}, 90(11):114032, 2014.

\bibitem{Hellerman:2014cba}
Simeon Hellerman, Shunsuke Maeda, Jonathan Maltz, and Ian Swanson.
\newblock {Effective String Theory Simplified}.
\newblock {\em JHEP}, 09:183, 2014.

\bibitem{Hellerman:2017upi}
Simeon Hellerman and Shunsuke Maeda.
\newblock {On Vertex Operators in Effective String Theory}.
\newblock 1 2017.

\bibitem{EliasMiro:2019kyf}
Joan Elias~Mir\'o, Andrea~L. Guerrieri, Aditya Hebbar, Jo\~ao Penedones, and Pedro Vieira.
\newblock {Flux Tube S-matrix Bootstrap}.
\newblock {\em Phys. Rev. Lett.}, 123(22):221602, 2019.

\bibitem{Caselle:2021eir}
Michele Caselle.
\newblock {Effective String Description of the Confining Flux Tube at Finite Temperature}.
\newblock {\em Universe}, 7(6):170, 2021.

\bibitem{EliasMiro:2021nul}
Joan Elias~Mir\'o and Andrea Guerrieri.
\newblock {Dual EFT bootstrap: QCD flux tubes}.
\newblock {\em JHEP}, 10:126, 2021.

\bibitem{Brandt:2016xsp}
Bastian~B. Brandt and Marco Meineri.
\newblock {Effective string description of confining flux tubes}.
\newblock {\em Int. J. Mod. Phys. A}, 31(22):1643001, 2016.

\bibitem{Teper:2009uf}
Michael Teper.
\newblock {Large N and confining flux tubes as strings - a view from the lattice}.
\newblock {\em Acta Phys. Polon. B}, 40:3249--3320, 2009.

\bibitem{Brandt:2009tc}
Bastian~B. Brandt and Pushan Majumdar.
\newblock {Spectrum of the QCD flux tube in 3d SU(2) lattice gauge theory}.
\newblock {\em Phys. Lett. B}, 682:253--258, 2009.

\bibitem{Athenodorou:2010cs}
Andreas Athenodorou, Barak Bringoltz, and Michael Teper.
\newblock {Closed flux tubes and their string description in D=3+1 SU(N) gauge theories}.
\newblock {\em JHEP}, 02:030, 2011.

\bibitem{Billo:2011fd}
Marco Billo, Michele Caselle, and Roberto Pellegrini.
\newblock {New numerical results and novel effective string predictions for Wilson loops}.
\newblock {\em JHEP}, 01:104, 2012.
\newblock [Erratum: JHEP 04, 097 (2013)].

\bibitem{Athenodorou:2011rx}
Andreas Athenodorou, Barak Bringoltz, and Michael Teper.
\newblock {Closed flux tubes and their string description in D=2+1 SU(N) gauge theories}.
\newblock {\em JHEP}, 05:042, 2011.

\bibitem{Brandt:2013eua}
Bastian~B. Brandt.
\newblock {Spectrum of the open QCD flux tube in $d = 2 + 1$ and its effective string description}.
\newblock {\em PoS}, EPS-HEP2013:540, 2013.

\bibitem{Caselle:2015tza}
Michele Caselle, Alessandro Nada, and Marco Panero.
\newblock {Hagedorn spectrum and thermodynamics of SU(2) and SU(3) Yang-Mills theories}.
\newblock {\em JHEP}, 07:143, 2015.
\newblock [Erratum: JHEP 11, 016 (2017)].

\bibitem{Athenodorou:2016kpd}
Andreas Athenodorou and Michael Teper.
\newblock {Closed flux tubes in D = 2 + 1 SU(N ) gauge theories: dynamics and effective string description}.
\newblock {\em JHEP}, 10:093, 2016.

\bibitem{Brandt:2017yzw}
Bastian~B. Brandt.
\newblock {Spectrum of the open QCD flux tube and its effective string description I: 3d static potential in SU(N = 2, 3)}.
\newblock {\em JHEP}, 07:008, 2017.

\bibitem{Athenodorou:2020ani}
Andreas Athenodorou and Michael Teper.
\newblock {The glueball spectrum of SU(3) gauge theory in 3 + 1 dimensions}.
\newblock {\em JHEP}, 11:172, 2020.

\bibitem{Caristo:2021tbk}
Fabrizio Caristo, Michele Caselle, Nicodemo Magnoli, Alessandro Nada, Marco Panero, and Antonio Smecca.
\newblock {Fine corrections in the effective string describing SU(2) Yang-Mills theory in three dimensions}.
\newblock {\em JHEP}, 03:115, 2022.

\bibitem{Luo:2023cjv}
Conghuan Luo, Andreas Athenodorou, Sergei Dubovsky, and Michael Teper.
\newblock {Confining strings and glueballs in $\mathbb{Z}_N$ gauge theories}.
\newblock {\em PoS}, LATTICE2023:375, 2024.

\bibitem{Zahn:2013yma}
Jochen Zahn.
\newblock {The excitation spectrum of rotating strings with masses at the ends}.
\newblock {\em JHEP}, 12:047, 2013.

\bibitem{Sonnenschein:2014jwa}
Jacob Sonnenschein and Dorin Weissman.
\newblock {Rotating strings confronting PDG mesons}.
\newblock {\em JHEP}, 08:013, 2014.

\bibitem{Helle}
Simeon Hellerman and Ian Swanson.
\newblock {String Theory of the Regge Intercept}.
\newblock {\em Phys. Rev. Lett.}, 114(11):111601, 2015.

\bibitem{Hellerman:2016hnf}
Simeon Hellerman and Ian Swanson.
\newblock {Boundary Operators in Effective String Theory}.
\newblock {\em JHEP}, 04:085, 2017.

\bibitem{Sever:2017ylk}
Amit Sever and Alexander Zhiboedov.
\newblock {On Fine Structure of Strings: The Universal Correction to the Veneziano Amplitude}.
\newblock {\em JHEP}, 06:054, 2018.

\bibitem{Sonnenschein:2018aqf}
Jacob Sonnenschein and Dorin Weissman.
\newblock {Quantizing the rotating string with massive endpoints}.
\newblock {\em JHEP}, 06:148, 2018.

\bibitem{Sonnenschein:2018fph}
Jacob Sonnenschein and Dorin Weissman.
\newblock {Excited mesons, baryons, glueballs and tetraquarks: Predictions of the Holography Inspired Stringy Hadron model}.
\newblock {\em Eur. Phys. J. C}, 79(4):326, 2019.

\bibitem{hooft2004minimal}
Gerard't Hooft.
\newblock Minimal strings for baryons.
\newblock {\em arXiv preprint hep-th/0408148}, 2004.

\bibitem{future}
Gabriel Cuomo, Sergei Dubovsky, Guzmán Hernández-Chifflet, Alexander Monin, and Shahrzad Zare.
\newblock To appear.
\newblock 2024.

\bibitem{Gaiotto:2017tne}
Davide Gaiotto, Zohar Komargodski, and Nathan Seiberg.
\newblock {Time-reversal breaking in QCD$_{4}$, walls, and dualities in 2 + 1 dimensions}.
\newblock {\em JHEP}, 01:110, 2018.

\bibitem{Artru:1974zn}
X.~Artru.
\newblock {String Model with Baryons: Topology, Classical Motion}.
\newblock {\em Nucl. Phys. B}, 85:442--460, 1975.

\bibitem{Rossi:1977cy}
G.~C. Rossi and G.~Veneziano.
\newblock {A Possible Description of Baryon Dynamics in Dual and Gauge Theories}.
\newblock {\em Nucl. Phys. B}, 123:507--545, 1977.

\bibitem{Kharzeev:1996sq}
D.~Kharzeev.
\newblock {Can gluons trace baryon number?}
\newblock {\em Phys. Lett. B}, 378:238--246, 1996.

\bibitem{Frenklakh:2023pwy}
David Frenklakh, Dmitri~E. Kharzeev, and Wenliang Li.
\newblock {Signatures of baryon junctions in semi-inclusive deep inelastic scattering}.
\newblock {\em Phys. Lett. B}, 853:138680, 2024.

\bibitem{Frenklakh:2024mgu}
David Frenklakh, Dmitri Kharzeev, Giancarlo Rossi, and Gabriele Veneziano.
\newblock {Baryon-number -flavor separation in the topological expansion of QCD}.
\newblock 5 2024.

\bibitem{altmann2024string}
Javira Altmann and Peter Skands.
\newblock String junctions revisited.
\newblock {\em arXiv preprint arXiv:2404.12040}, 2024.

\bibitem{Rossi:2016szw}
Giancarlo Rossi and Gabriele Veneziano.
\newblock {The string-junction picture of multiquark states: an update}.
\newblock {\em JHEP}, 06:041, 2016.

\bibitem{PhysRevD.95.034011}
Marek Karliner, Shmuel Nussinov, and Jonathan~L. Rosner.
\newblock $qq\overline{Q}\overline{Q}$ states: Masses, production, and decays.
\newblock {\em Phys. Rev. D}, 95:034011, Feb 2017.

\bibitem{Karliner:2020vsi}
Marek Karliner and Jonathan~L. Rosner.
\newblock {First exotic hadron with open heavy flavor: $cs\bar u\bar d$ tetraquark}.
\newblock {\em Phys. Rev. D}, 102(9):094016, 2020.

\bibitem{Takahashi:2000te}
Toru~T. Takahashi, H.~Matsufuru, Y.~Nemoto, and H.~Suganuma.
\newblock {The Three quark potential in the SU(3) lattice QCD}.
\newblock {\em Phys. Rev. Lett.}, 86:18--21, 2001.

\bibitem{Takahashi:2002bw}
Toru~T. Takahashi, H.~Suganuma, Y.~Nemoto, and H.~Matsufuru.
\newblock {Detailed analysis of the three quark potential in SU(3) lattice QCD}.
\newblock {\em Phys. Rev. D}, 65:114509, 2002.

\bibitem{Alexandrou:2002sn}
C.~Alexandrou, P.~de~Forcrand, and Oliver Jahn.
\newblock {The Ground state of three quarks}.
\newblock {\em Nucl. Phys. B Proc. Suppl.}, 119:667--669, 2003.

\bibitem{Bissey:2005sk}
F.~Bissey, F-G. Cao, A.~Kitson, B.~G. Lasscock, D.~B. Leinweber, A.~I. Signal, A.~G. Williams, and J.~M. Zanotti.
\newblock {Gluon field distribution in baryons}.
\newblock {\em Nucl. Phys. B Proc. Suppl.}, 141:22--25, 2005.

\bibitem{Bissey:2006bz}
F.~Bissey, F-G. Cao, A.~R. Kitson, A.~I. Signal, D.~B. Leinweber, B.~G. Lasscock, and A.~G. Williams.
\newblock {Gluon flux-tube distribution and linear confinement in baryons}.
\newblock {\em Phys. Rev. D}, 76:114512, 2007.

\bibitem{jahn2004baryons}
Oliver Jahn and Philippe De~Forcrand.
\newblock Baryons and confining strings.
\newblock {\em Nuclear Physics B-Proceedings Supplements}, 129:700--702, 2004.

\bibitem{Pfeuffer:2008mz}
Melanie Pfeuffer, Gunnar~S. Bali, and Marco Panero.
\newblock {Fluctuations of the baryonic flux-tube junction from effective string theory}.
\newblock {\em Phys. Rev. D}, 79:025022, 2009.

\bibitem{low2002spontaneously}
Ian Low and Aneesh~V Manohar.
\newblock Spontaneously broken spacetime symmetries and goldstone’s theorem.
\newblock {\em Physical review letters}, 88(10):101602, 2002.

\bibitem{nambu1970duality}
Yoichiro Nambu.
\newblock Duality and hydrodynamics.
\newblock {\em Broken Symmetry}, pages 280--301, 1970.

\bibitem{Goto:1971ce}
Tetsuo Goto.
\newblock {Relativistic quantum mechanics of one-dimensional mechanical continuum and subsidiary condition of dual resonance model}.
\newblock {\em Prog. Theor. Phys.}, 46:1560--1569, 1971.

\bibitem{Caselle:2013dra}
Michele Caselle, Davide Fioravanti, Ferdinando Gliozzi, and Roberto Tateo.
\newblock {Quantisation of the effective string with TBA}.
\newblock {\em JHEP}, 07:071, 2013.

\bibitem{Dubovsky:2014fma}
Sergei Dubovsky, Raphael Flauger, and Victor Gorbenko.
\newblock {Flux Tube Spectra from Approximate Integrability at Low Energies}.
\newblock {\em J. Exp. Theor. Phys.}, 120:399--422, 2015.

\bibitem{Chen:2018keo}
Chang Chen, Peter Conkey, Sergei Dubovsky, and Guzm\'an Hern\'andez-Chifflet.
\newblock {Undressing Confining Flux Tubes with $T\bar T$}.
\newblock {\em Phys. Rev. D}, 98(11):114024, 2018.

\bibitem{Note1}
In the $SU(3)$ Yang-Mills theory, $\Omega $ should be understood as the trace of a Wilson line in the fundamental representation.

\bibitem{ishibashi1989boundary}
Nobuyuki Ishibashi.
\newblock The boundary and crosscap states in conformal field theories.
\newblock {\em Modern Physics Letters A}, 4(03):251--264, 1989.

\bibitem{andreev2016some}
Oleg Andreev.
\newblock Some aspects of three-quark potentials.
\newblock {\em Physical Review D}, 93(10):105014, 2016.

\bibitem{Imamura:2004tf}
Yosuke Imamura.
\newblock {On string junctions in supersymmetric gauge theories}.
\newblock {\em Prog. Theor. Phys.}, 112:1061--1086, 2004.

\bibitem{Note2}
The analogous configuration in large-$N$ theories is where three $\protect \frac {N}{3}$-composite strings are connected at the junction. The composite string tension as well as the junction mass scale as $O(N)$ \cite {Witten:1979kh}, while the EFT cutoff depends on the fundamental string tension that scales as $O(N^0)$. In the large-$N$ limit, our discussion applies to composite string tension and junction mass in units of $N$, and the fact that the composite string tension does not set the cutoff scale indicates no perturbative instability.

\bibitem{teixeira2007energy}
PIC Teixeira and MA~Fortes.
\newblock Energy and tension of films and plateau borders in a foam.
\newblock {\em Colloids and Surfaces A: Physicochemical and Engineering Aspects}, 309(1-3):3--6, 2007.

\bibitem{dymarsky20183d}
Anatoly Dymarsky, Filip Kos, Petr Kravchuk, David Poland, and David Simmons-Duffin.
\newblock The 3d stress-tensor bootstrap.
\newblock {\em Journal of High Energy Physics}, 2018(2):1--48, 2018.

\bibitem{Dymarsky:2013wla}
Anatoly Dymarsky.
\newblock {On the four-point function of the stress-energy tensors in a CFT}.
\newblock {\em JHEP}, 10:075, 2015.

\bibitem{Note3}
For critical strings, the cubic couplings were recently discussed in~\cite {rosenhaus2022chaos}.

\bibitem{Bonati:2021vbc}
Claudio Bonati, Michele Caselle, and Silvia Morlacchi.
\newblock {The Unreasonable effectiveness of effective string theory: The case of the 3D SU(2) Higgs model}.
\newblock {\em Phys. Rev. D}, 104(5):054501, 2021.

\bibitem{Witten:1979kh}
Edward Witten.
\newblock {Baryons in the 1/n Expansion}.
\newblock {\em Nucl. Phys. B}, 160:57--115, 1979.

\bibitem{rosenhaus2022chaos}
Vladimir Rosenhaus.
\newblock Chaos in a many-string scattering amplitude.
\newblock {\em Physical Review Letters}, 129(3):031601, 2022.

\end{thebibliography}

\end{document}